\documentclass[runningheads]{llncs}

\usepackage{lineno}
\usepackage{float}
\usepackage{graphicx}
\usepackage{hyperref}
\usepackage{booktabs}
\usepackage[tight]{subfigure}
\usepackage{caption}

\usepackage{afterpage}
\usepackage{graphicx}

\usepackage{amsmath,amssymb,amsfonts}
\usepackage{algorithmic}
\usepackage{textcomp}
\usepackage[natural,table]{xcolor}
\usepackage{tkz-euclide}
\usepackage{framed}
\usepackage{tikz}
\usepackage{adjustbox}
\usepackage{url}
\usepackage{libertine}
\usepackage{pdflscape}
\usepackage[xspace]{ellipsis}

\usetikzlibrary{positioning, arrows, shapes,chains}
\usepackage{color}
\usepackage{transparent}
\graphicspath{img/}
\usepackage{cleveref}

\Crefformat{figure}{Figure~#1}
\Crefmultiformat{section}{Sections~#2#1#3}{ and~#2#1#3}{, #2#1#3}{ and~#2#1#3}

\usepackage[
key=WieseBrandtvanHoorn2023,
  year=2023,
  publication={17th European Conference on Software Architecture, {ECSA} 2023},
  nocopyright,
  nourl,
  nobib,
]{authorarchive}

\begin{document}
\title{Learning From Each Other: How Are Architectural Mistakes Communicated in Industry?}
\titlerunning{How Are Architectural Mistakes Communicated in Industry?}

\author{Marion Wiese
\orcidID{0000-0003-0160-9531} \and
Axel-Frederik Brand 
\and \\
André van Hoorn
\orcidID{0000-0003-2567-6077}}
\authorrunning{M. Wiese et al.}
\institute{Universität Hamburg, Department of Informatics, Hamburg, Germany  
\email{marion.wiese@uni-hamburg.de}
}

\maketitle              % typeset the header of the contribution

\begin{abstract}
% Context: The importance of the research questions addressed by the review. 
% 1. A general statement introducing the broad research area of the particular topic being investigated.
\textit{Context.} Own experiences and faulty decisions can be an important source of information for software architects. The experiences and mistakes of other architects can also be valuable information sources. %Architects can also profit from experiences and mistakes of other architects.
% Objectives: The questions addressed by the systematic review. 
% 2. An explanation of the specific problem (difficulty, obstacle, challenge) to be solved.
% 3. A review of existing or standard solutions to this problem and their limitations. 
\textit{Goal.} Under the assumption that the knowledge about faulty decisions, i.e., mistakes, regarding software architecture is not shared adequately in practice, this work qualitatively investigates the handling and particularly communication of those mistakes by software architects. 
% Methods:	Data Sources, Study selection, Quality Assessment and Data extraction. 
\textit{Method.} We conducted a grounded-theory study in which we interviewed ten German software architects from various domains. 
% Results: Main finding including any meta-analysis results and sensitivity analyses. 
% 4. An outline of the proposed new solution. 
\textit{Results.} We identified software architects' definitions of architectural mistakes, their handling of these mistakes, and their preferred communication strategies regarding these mistakes. 
We found that architects communicate mistakes mainly within their project teams and seldom within or across companies.  
% Conclusions: Implications for practice and future research.
% 5. A summary of how the solution was evaluated and what the outcomes of the evaluation were.
\textit{Conclusions.} 
We derived strategies to make learning and prevention of mistakes more effective. 
To share experiences and knowledge beyond architects' peer groups, companies should invest more effort in discussing mistakes more consciously and create an environment where mistakes can be discussed openly. 

\end{abstract}

    \section{Introduction}
    \label{sec:Introduction}
     
        %Introducing SA and SA mistakes / judgement errors
    
            ``Errare humanum est''\,---\,to err is human, is common knowledge since ancient\linebreak times~\cite{hieronymus_errare_nodate}. 
            In software architecture, we also make mistakes. 
            By sharing our experiences with mistakes, we enhance the chance of not repeating them.  
            In other domains, e.g., economics, it has become common to learn from each other's mistakes by sharing experiences, e.g., by conducting ``Fuck up nights''~\cite{noauthor_fuckup_2023}.
            However, in software architecture (SA), it seems that mistakes are not a commonly discussed topic.
            Even using the term ``mistake'' in the context of SA may seem offensive.
            
            Technical debts are constructs in software systems that are beneficial in the short term but hinder changes in the long term~\cite{Avgeriou2016a}. 
            Technical debt in SA was identified as one of the most dangerous types~\cite{Martini2015a} because there is a lack of refactoring strategies~\cite{Besker2018a}. 
            % %CUT 
            % Fowler created a two-by-two matrix to explain the causes of technical debt by differentiating between deliberate/inadvertent and reckless/prudent technical debt~\cite{Fowler2009}. 
            % Fowler explained the prudent and inadvertent quadrant with the sentence: ``Now we know how we should have done it.'' 
            % This is the essence of making and detecting mistakes. 
            % Architectural technical debt often falls into the prudent and inadvertent quadrant, which reflects in research on uncertainty in architectural decision-making~\cite{sobhy_evaluation_2021}. 
            % Architects make prudent and deliberate choices, but these choices turn out to inadvertently be technical debt later when more knowledge and experience are collected. 
            % %bis hier
            
            One goal to avoid architectural technical debt is to amass knowledge and have at least some experience before deciding on an SA issue. 
            A lot of research focuses on the optimization of SA processes, particularly with regard to SA decision-making~\cite{Bhat2020,tang_human_2017} and gathering SA knowledge~\cite{weinreich_software_2016}. %~\cite{ali_babar_software_2009}.
            % In our previous work we
            Soliman et al.\ found that when searching for SA knowledge on Google, experience reports on design decisions and drawbacks of solution options are relevant to a decision but are underrepresented in search results~\cite{Soliman2021c}. 
            
        %Rationale
            Architects deal with mistakes, i.e., design decisions that turn out to be sub-optimal, by doing ``post-mortems''~\cite{bjornson_improving_2009} %~\cite{dingsoyr_augmenting_2001}
            or by defining anti-patterns~\cite{richards_software_2015}, and architecture\linebreak smells~\cite{mumtaz_systematic_2021}. %~\cite{baabad_software_2020}
            %CUT  (Mit Referenz!)
            %, and sometimes related refactorings~\cite{stal_chapter_2014}.  
            % bis hier
            However, the question of how to communicate the mistakes between architects is not yet researched systematically to the best of the authors' knowledge. 
            
            %Goal + Method
            Our goal is to fill this gap by conducting a grounded theory study. 
            We interviewed ten software architects to gather their approaches to sharing bad experiences regarding decisions in SA, i.e., SA mistakes (SAMs). 
        %    
        % RQs
            To follow this goal, we answer the following research questions~(RQs):
            \begin{itemize}
                
                \item %[] 
                \textbf{RQ 1: \textit{How do architects define a SA mistake?}}
                    This question lays the foundation for our work, as the term mistake is not yet commonly used or defined in relation to SA.
                    
                \item %[] 
                \textbf{RQ 2: \textit{How do architects manage SA mistakes?}}
                    We explain how the architects identify and handle SAMs. 
                    %Parts of these insights might not be novel. 
                    %However, 
                    It is relevant to understand the overall management of SAMs to identify communication within these processes. 
                    Furthermore, we created a comprehensive model of the SAMs' management. %CUT:, which is an overview of how architects deal with mistakes in practice.
                    
                \item %[] 
                \textbf{RQ 3: \textit{How do architects communicate SA mistakes?}}
                    We identified the ways architects communicate SAMs and particularly on which level, i.e., personal, team-wide, company-wide, or even outside of their company. 
                    Furthermore, we identified factors that suppress or promote the communication of SAMs. 
                    Thereby, we uncovered opportunities for process improvements. 
                    
            \end{itemize}
        
        %Contribution - general
            \noindent As the contribution of our study, we answer the research questions and provide:
            \begin{itemize}
                \item A definition of SA mistakes
                \item A theory of a model of SA mistakes' management
                \item SA mistakes communication issues and potential for improvement
            \end{itemize}
            
        % study outline
            
            %In the following section, %we present the background information that forms the basis for motivating and understanding this study.
            In \Cref{sec:Method}, 
            we explain our research methodology. 
            We present the results and answer the research questions in \Cref{sec:Results}.  
            In \Cref{sec:Discussion}, we discuss our findings.  
            We compare our study with related work in \Cref{sec:RelatedWork} and discuss the threats to validity of our study in \Cref{sec:ThreatsToValidity}. 
            With \Cref{sec:Conclusion}, we conclude our paper and summarize the contributions for practitioners and researchers. 
            Additional material is available~\cite{AdditionalMaterial2023}.
    
    \section{METHOD}
    \label{sec:Method}
        To answer our research question, we did a grounded theory study.  
        Grounded Theory was first introduced by Glaser \& Strauss~\cite{glaser_discovery_1968}. 
        In grounded theory, the collected data is analyzed interpretatively, and identified concepts are comprehensibly collected.  
        The concepts are compared, structured, and summarized into categories. 
        From these categories, theses are derived that are valid across the data.  
        The final goal of Grounded Theory is to develop a theory on the researched topic. This theory must then be validated by further studies, e.g., surveys.
        %By means of analytical memos, the thoughts of the researchers are recorded during the analysis process.  
        The coding and analysis process takes place in parallel to the collection of new data, whereby the categories are iteratively reconsidered, compared, and questioned.  
        This is referred to as the method of constant comparison. 
        The process of collecting and analyzing data is carried out until the theory is saturated. 
        Saturation is reached when new data hardly generate any new insights in the form of new concepts and categories. 

        %CUT:To satisfy this grounded theory process, we 
        We collected and analyzed data in three phases that merged into one another. 
        In each phase, we adapted the interview guide, did the interviews, and analyzed the corresponding data.  
        %
        %% Kommt denn am Ende ein Theorie raus? Ansosnten vielleicht diesen Teil  leiber einfach weglassen?
        %CUT: For our study, we mainly used the constructivist methodology according to Charmaz~\cite{charmaz_constructing_2014}. 
        %CUT: Constructivism in this context means that Charmaz sees the role of the researcher as constructing a theory rather than being a discovering or interpretive character. 

        % The incorporation of prior knowledge is allowed in both methods and strengthens the later collection of data. 
        % In both methods, the research question can develop further in the course of the research.
        % The main differences between the methodologies lie in the process of coding and analysis (Sebastian, 2019), which will be discussed explicitly in the following sections.

        \subsection{Data collection}
        \label{sec:data_collection}        
        We conducted ten interviews with architects selected according to our sampling strategy and in accordance with our interview guide. We transcribed the recordings of the interviews for further analysis. 
        All interviews were held in the German language.
        
        \paragraph{\textbf{Sample.}}
        \label{sec:sampling_strategy}
        We recruited software architects from our personal network, via social media and speakers from conferences in three phases. 
        All participants completed a survey about their personal background, e.g., domain, company size, and experience level (see~\cite{AdditionalMaterial2023}). 
        With this information and the iterative approach, we ensured to gather participants from different domains, with different levels of experience, and consulting and company architects.
        A variety of participants is important to validate that code saturation is reached for the given context, i.e., German software architects.
        %All participants consented to the use of the interview recordings.      
        An overview of the participants can be seen in \Cref{tab:participants}. %TODo
        
        \begin{table}
            \centering
            \setlength{\tabcolsep}{0.2cm}
            \rowcolors{2}{gray!25}{white!50}
            \footnotesize
           	\begin{tabular} {llllrrc} %%{p{1cm} p{3cm} p{3cm} p{2m} p{2cm} p{1cm} } %
        		\toprule
           	   	\textit{Participant} & \textit{Company}  & \textit{Company size}  & \textit{Team} & \textit{Years} & \textit{Years as} & \textit{Gender} \\ 
           	   	  &  \textit{Domain} &  \textit{(\#employees)} & \textit{organization} & \textit{in IT} & \textit{architect} &  \\ 
        		%\midrule 
        		P1 	& Media           & 1000-5000 & agile & $>$20 & 5-10  &  f 	        \\ 
        		P2 	& Consulting      & $>$5000   & agile & 10-15 & 10-15 &  m   	    \\ 
        		P3 	& Consulting      & $>$5000   & agile & $>$20 & 10-15 &  m  	    \\ 
        		P4 	& Aviation        & $>$5000   & agile & 15-20 & 10-15 &  m   	    \\ 
        		P5 	& Industry        & $>$5000   & mix (agile/classic) & 10-15 & 2-5   &  m   	    \\ 
        		P6 	& Consulting      & 0-100     & agile & 15-20 & 5-10  &  m   	    \\ 
        		P7 	& Consulting      & 100-1000  & agile & 5-10  & 5-10  &  m   	    \\ 
        		P8 	& Consulting      & 1000-5000 & agile & $>$20 & 15-20 &  m   	    \\ 
        		P9 	& Consulting      & 0-100     & agile & $>$20 & $>$20 &  m   	    \\ 
        		P10	& Insurance       & $>$5000   & agile & $>$20 & $>$20 &  f  	    \\ 
           % 			\midrule 
           % 			sum 	&  & & & & & & 6 & 11 & 17 & 17 & 16 & 2\\ 
        			\bottomrule
   		\end{tabular}
   		\caption{Study Participants}
            \label{tab:participants}
        \end{table}
        \vspace{-10mm}
        % The guideline comprised questions regarding:
        % \begin{itemize}
        %     \item Introductory question with reference to given definitions of software architecture 
        %     \item Introductory questions related to SAMs and searching for examples
        %     \begin{itemize}
        %         \item Questions on probing one example of a SAM
        %     \end{itemize}
        %     \item Main questions on the communication of SAMs
        %     \begin{itemize}
        %         \item Classification of the role of communication and how the person communicates
        %         \item Influence of ``mistake culture'' 
        %     \end{itemize}
        %     \item Retrospective or classifying exit questions
        % \end{itemize}
        
        \paragraph{\textbf{Interview Guideline.}}
        \label{sec:interview_guideline}
            We adapted the guideline before each of the three phases to iteratively focus on the emerging theory. All guidelines are available on Zenodo~\cite{AdditionalMaterial2023}
            In the end, i.e., in phase three, this led to the following nine questions:
            \begin{enumerate}
                %\item We presented the definition of SA according to ISO 42010 
                \item What is a SAM or faulty decision in this context? 
                \item Can you give me examples from your own experience of a SAM or faulty decision? (plus sub-questions regarding the handling of this specific example)
                \item What is the role of communicating bad decisions and mistakes for software architects? 
                \item How do you talk about bad decisions with your colleagues or other architects? 
                \item How do you talk about SAMs with your team lead or other managers? Does your communication style differ from your colleagues?
                \item How do different error cultures (e.g. through in-sourcing / out-sourcing or international work / different companies) influence you in dealing with SAMs or faulty decisions?
                \item How are faulty decisions or SAMs dealt with in your company for the purpose of knowledge transfer?
                \item How has the way you deal with and communicate mistakes changed over time? 
                \item With regard to concrete faulty decisions\,---\,how and with whom should software architects communicate them? 
            \end{enumerate}

        \subsection{Data Analysis}
        \label{sec:data_analysis}
        We coded the transcriptions according to Charmaz's guidelines and reached theoretical saturation after ten interviews~\cite{charmaz_constructing_2014}.
        
       \paragraph{\textbf{Coding.}}
        \label{sec:coding}
            
            According to Charmaz~\cite{charmaz_constructing_2014}, the process of coding consists of two phases: initial coding and focused coding. 
            \textit{Initial coding} is an inductive approach. 
            Data segments are considered word-by-word and sentence-by-sentence, and all initial analytical concepts are collected without further comparison. 
            \textit{Focused coding} describes hierarchical sorting and abstracting, i.e., merging of concepts or categories as well as the selection of core categories on which the analysis is focused. 
            %In practice, the categories are compared with each other, and the number of categories is reduced. 
            The process no longer focuses on the raw data but only on the codes. 
    
            Furthermore, through cross-coding phases, during which we constantly compared the codes and coded segments, we iterated through every interview multiple times to make sure that the codes were comparable and identical in meaning (Code list see~\cite{AdditionalMaterial2023}). 
            %A code list is part of our additional material (see~\cite{AdditionalMaterial2023}). 

       \paragraph{\textbf{Theoretical Saturation.}}
        \label{sec:saturation}
            The goal of the methodology is to achieve theoretical saturation.  
            Theoretical saturation means that the evaluation of new data, for example, through further interviews, no longer fundamentally changes the existing structure of the category system. 
            To evaluate saturation, the number of new codes in each new interview is evaluated, and the goal is to amass fewer codes with each interview's evaluation, particularly focus codes.
            
            As we used cross-coding, the number of codes varied a lot and even got smaller if codes were merged in the process.
            During our last three interviews, we reduced the codes from 291~codes to 264~codes. 
            During the last interview, we still gained 13~codes (see~\cite{AdditionalMaterial2023}). 
            However, all these new codes were initial codes and did not provide new insides for the report of the results.
            %initial codes (see~\Cref{sec:data_analysis}), and we only gained one focus code (see~\Cref{sec:data_analysis}) during the last three interviews.
            Therefore, we assume theoretical saturation for the results presented in this paper. % on the hierarchy level that we report here.

    \section{Results}
    \label{sec:Results}
    
    In this section, we %present the results of our research questions.
    %We 
    give a detailed overview of the definition and characteristics of SA mistakes and how to detect them. 
    We present how architects handle detected SAMs and, finally, summarize how they communicate the detected SAMs. Each section concludes with a summary.

        %----------------------RQ1 - Discussion ------------------------
        \subsection{RQ1: Definition and Characteristics of SA Mistakes}
        \label{sec:Results-RQ1}
        From the transcripts, we extracted the following definition of SAMs and their characteristics. 
        \paragraph{\textbf{Definition.}}
        A SA decision is classified as a mistake if, for a given problem, an alternative solution exists that solves the problem better than the implemented solution. 
        The choice of an architect is always associated with trade-offs and considerations. 
        Whether a problem is solved ``better'' is not easy to judge objectively. 
        Therefore, some participants describe SAMs by their negative impact on non-functional quality attributes. 
        % A SAM can be identified if several alternatives exist for the intended use and an inappropriate choice is made. 
        % However, this assessment is not perceived as a simple one by the participants.
        \textit{``When it comes to architectural decisions, that [a mistake] is if something doesn't work, I mean more the non-functional [if] you will not meet a non-functional requirement.''} %CUT:Either from a priori or after the fact. So maybe not yet at the moment you built it, but with a slight shift in a requirement or another context of use.''} 

        The effects of a SAM are often noticed particularly late, i.e., they have a long feedback cycle. 
        \textit{``The question is, however, after some time, when you reach an appropriate scale, and the application has a corresponding scope, whether you then realize: Yes, you've got it working, but it's not really running ideally.''}

        SAMs are considered difficult to identify because the alternatives are not directly known to the architects. 
        Even if the architects know the alternatives, % these are difficult to assess. SA is complex, and
        the effects of SA decisions are difficult to assess. 
        \textit{``However, you often do not see what an alternative implementation would have brought and what difficulties it would have brought with it.''} 
        %CUT:
        %You only ever see the difficulties of the thing you've built. Everything else always looks as if it were quite simple.''} 

        SA is an overarching topic because SA should fulfill many different, conflicting goals with many influencing factors and constraints. 
        Therefore, problems are not simply attributable to individual decisions:
        \textit{``So I think with architecture mistakes, it's often the case that you can't even say that they did something wrong at that point, but it is rather that somehow the realization arises that something is not going right.''}

        \vspace{0,2cm}
        \noindent
        \fbox{%
            \parbox{0.97\textwidth}{%
            The goal of SA is to meet non-functional requirements. 
            A SAM is defined as the non-fulfillment of these requirements.
            SA pursues many different and contradictory goals, and alternatives are often not obvious, which makes the evaluation difficult. 
            A wrong decision often becomes apparent only after a long time.
            }
        }

        \paragraph{\textbf{Characteristics.}}
        For the participants, a SAM results from the comparison of an actual state with a target state. 
        One participant explained that a SAM is
        \textit{``everything that turns out afterward that it somehow doesn't work out the way you imagined it.''}
        This comparison between the actual and target state %CUT 
      %  represents the core of the defect definition, and %bis hier
        is a subjective evaluation. 
        %\textit{``Mistake is a difficult concept. A mistake is also an evaluation. There is no abstract form of a 'mistake'. First of all, every decision is just there.''}
        The subjective approach results in subjective error levels.  
        The situations are classified as acceptable or not acceptable, which in situations with several actors can result in a different understanding of the situation.  
        \textit{``We use our own developed framework, which is already twelve years old because Person C wants it that way, and he gets in the way when we use something else. \ldots these are not only wrong decisions, this is almost wanton destruction.''.}
        
        The term ``mistake'' is subjectively perceived in different ways, i.e., there are different connotations associated with the term.
        Some architects mentioned that making SAMs is normal. 
        \textit{``I think making faulty decisions is okay. It can happen, and you have to deal with that somehow.''} %CUT: So, the expectation that every decision that I make will be mistake-free in retrospect -- I don't think that can be expected.''} 
        Some participants also mentioned the positive aspect of making SAMs and learning from them. 
        \textit{``It's that feeling that to do something really well, you have to have failed at it at least once.''}
        % %CUT:
        % That's an important point that you've been in projects at least once or twice, where you've realized that you've really gotten into the shit, and you have to dig yourself out of the mud again.''} 
        However, much more frequently, the participants mention negative aspects of making SAMs: %CUT: \textit{``I rarely use mistakes as a word when talking to customers and colleagues, I think\ldots. Of course, it is a mistake, in a way. But it always has a negative connotation.''} 
        \textit{``However, it is unfortunately often the case that due to the fact that SA [decisions] are the expensive decisions, mistakes very often cannot be communicated \ldots because then it often looks as if people have wasted a lot of money out of stupidity.''} 

        Another characteristic of mistakes is that they are avoidable. 
        We already established that architects think ``making mistakes is normal''. 
        This can be interpreted as ``making mistakes is unavoidable''. 
        Yet, this inevitability only applies to mistakes in general.  
        \textit{``I don't think we could have actually done much differently about the situation. \ldots That falls under [can be considered as] `shit happens.'''} 
        However, on the level of a single mistake, mistakes are considered to be avoidable. 
        \textit{``If you had trusted the expertise [of the people involved], this wouldn't have happened.''} 

         Finally, another characteristic of a SAM is that we often don't know whether a decision is a mistake until a certain amount of time has passed.
        \textit{``I think, overall, it came to light too late that this was even a problem at all.''} 

        \vspace{0,2cm}
        \noindent
        \fbox{%
            \parbox{0.97\textwidth}{%
            \begin{itemize} [leftmargin=*, topsep=0pt, itemsep=0pt]
                %\item SAMs are situations that deviate from the ideal that is actually desired.
                \item A SAM is a subjective evaluation of a situation. 
                \item A SAM may be viewed as acceptable by some and unacceptable by other persons.
                \item The term SAM can be given positive or negative connotations.
                \item The single SAM can be seen as avoidable.
                \item In general, the occurrence of SAMs is regarded as unavoidable.
                \item A SAM is often recognized retrospectively after a certain period of time.
            \end{itemize}
            }
        }

        \subsection{RQ2: Management of SA Mistakes}
        \label{sec:Results-RQ2}
\begin{figure}%[H]
            \centering
            \includegraphics[width=0.85\textwidth]{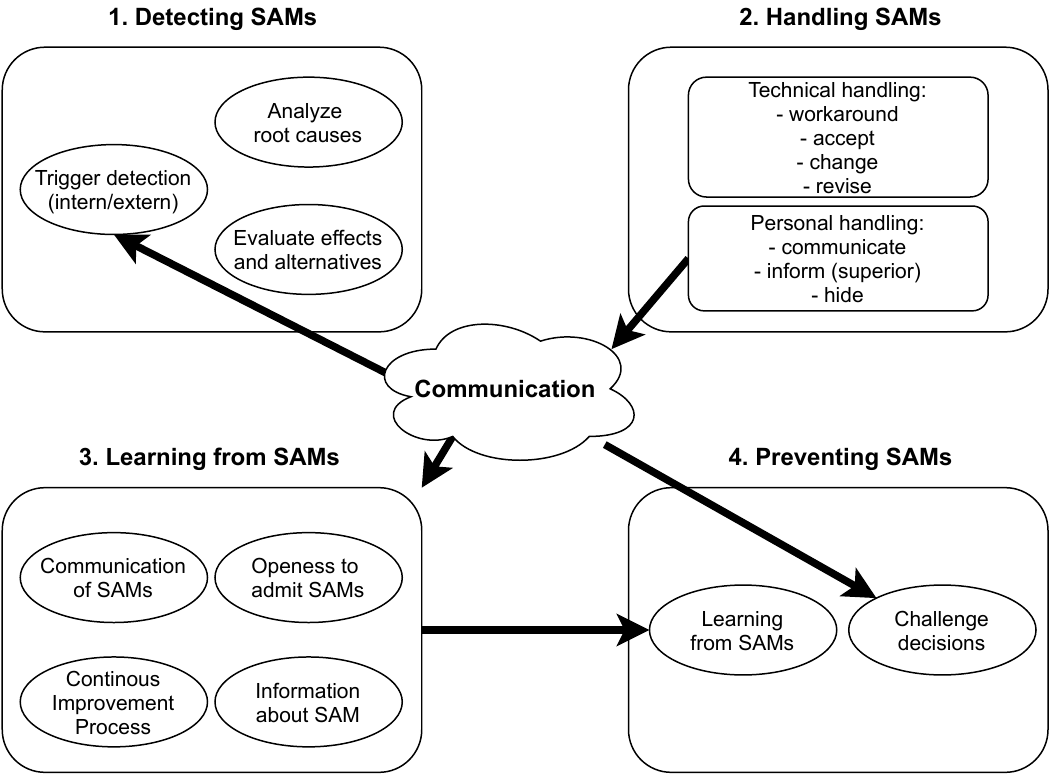}
            \caption{Theory of a model of SAM's Management}
            \label{fig:SA_Mistake_Management}
            %\vspace{-5mm}
        \end{figure}

        We asked the participants questions about their example situation, particularly how and when they came to the conclusion that the situation was a SAM and how the situation was handled afterward. 
        The results are clustered into four subsequent actions:
        \begin{enumerate} [leftmargin=25pt, topsep=1pt, itemsep=0pt]
            \item Detecting SAMs
            \item Handling SAMs
            \item Learning from SAMs
            \item Preventing SAMs
        \end{enumerate}
        We created a theory of a model to visualize the communication-based inter-dependencies between these categories and their sub-categories, depicted in \Cref{fig:SA_Mistake_Management}. 

        \paragraph{\textbf{Detecting SAMs.}}
        Mistake detection is typically initiated by a trigger.  
        These triggers can be driven by the architects themselves, e.g., dissatisfaction with the situation or questioning of requirements. 
        \textit{``%At some point, there was a moment for me when I realized: 
        If I look at it [the decision] like this, to what extent is this decision that was made in management supported by the teams that use it [the system] \ldots? %, i.e., by sales or product management? 
        There was a moment when I realized: Not at all.''} 
        More often, however, triggers from outside were mentioned, e.g., changes in project team structure, functional tests with users, changes in technical conditions, architecture reviews, or new requirements. 
        \textit{``So you suddenly have certain load situations and realize that you have now found the bottleneck. 
        You made a decision back then that now presents you with extreme problems.''} 

        In the next step, the trigger is used to try to understand the situation. 
        It is not always apparent whether there is a mistake or exactly where the mistake is. 
        \textit{``If you notice a problem somewhere and you ask the swarm [all colleagues combined], then the first question is: Where does it come from? What is the cause? 
        And then you come to the conclusion that at some point you made a wrong decision or a wrong assessment.''}
        
        To evaluate the situation, the effects of the SAM are analyzed, alternatives are examined, and, if possible, reasons for decisions are reconciled. 
        %The evaluation of the situation influences the resulting follow-up action. 
        \textit{``Often the question is: Was this the best, the second best, or the fourth best solution for something? 
        Nevertheless, the question is: Does it work? Or does it not?''}
        %And if it does, then it is probably more sensible to live with it than to switch everything to the best solution, come hell or high water.“} 
        
        \vspace{0,2cm}
        \noindent
        \fbox{%
            \parbox{0.97\textwidth}{%
            A SAM is detected when a trigger causes the re-evaluation of a situation. 
            These triggers can originate from the architects themselves (internal) but also from external triggers such as changes in the project team, new requirements, functional tests, or external audits. 
            After detection, the situation is analyzed, and alternatives are examined, which triggers a follow-up action.
            }
        }

        \paragraph{\textbf{Handling SAMs.}}
        The SAM's handling describes the follow-up action resulting from the evaluation.
        This corresponds to the technical handling of the SAMs.  
        Other actions are directly performed by the people involved, which corresponds to the personal SAM handling.
        The core of SAM handling is often the intention to do something different in the future. 
        \textit{``Well, I wouldn't do it that way again. I would try it differently.''} 
        %Whether or not it's really a 100 percent bad decision, I can't tell you that either.''} 

        Regarding technical handling, sometimes a workaround can be created to restore functionality.
        The most common handling approaches are to accept a sub-optimal situation or to develop the architecture further.  
        %\textit{``We now have a situation where we know in principle that the architecture is not optimal.  
        %So that it could be easier, and we also produce maintenance costs with it. 
        %But we still don't change it because the investment to change it would simply be too high.''} 
        \textit{``And if it [the SA] does [work], then it is probably more sensible to live with it than to switch everything to the best solution, come hell or high water.''} 
        Only rarely do architects revise their architecture completely. 
        \textit{``This project completely failed. We came in there as a different consulting company and had to stop everything, redo everything, and set everything up again.''} 

        As personal SAM handling, all participants report that as a result of a SAM's assessment, the SAM must be communicated, e.g., \textit{``We have highlighted these problems over and over again. We have said this again and again.''} 
        Furthermore, it is important to communicate with superiors to get their support. 
        \textit{``I talk very openly with my direct superiors if I think something is right or wrong. 
        Then we usually think about whether we want to change it or not.''} 
        Interestingly, many participants reported that they did not use to communicate SAMs in the past, and this changed over time. 
        \textit{``In the past, I was much more defensive about admitting my own mistakes.''} 
        % %CUT:
        % I once had a phase when I believed I made no mistakes. 
        % I think everyone has had that at one time or another. 
        % And when you have made mistakes, then you have vigorously denied that it was a mistake or that it was not possible to decide differently. 
        % %bis hier
        % That actually developed over time. 
        % %oder hier
        %In the past, I was much more defensive about admitting my own mistakes.''}       
        The topic of SAMs' communication will be studied in more detail in the following~\Cref{sec:Results-RQ3}.

% %CUT: Ist interessant, wird aber im Zweifel nicht vermisst
%         Communication can also be important in limiting the impact of SAMs. 
%         Due to the lack of communication in the following situation, the SAM only became visible after the system failure.  
%         \textit{``[I]t came to light too late\ldots 
%         No one came and said: 'My database is getting fuller, or the system is slowing down. 
%         They may have thought that, but they accepted it.''} 

        \vspace{0,2cm}
        \noindent
        \fbox{%
            \parbox{0.97\textwidth}{%
            SAMs are handled on a technical and personal level.
            The technical handling comprises creating workarounds and accepting, changing, or revising the SA.
            Often the sub-optimal solution is accepted with the intention of deciding differently next time.  
            Personal handling comprises communication with colleagues and superiors.
            The participants conceded their openness to communicate SAMs did develop over time. 
            }
        }

        \paragraph{\textbf{Learning from SAMs.}}
        We asked the architects if and how they learn from SAMs, i.e., we asked them about the requirements to learn from SAMs. 

        Firstly, all participants consider that communication is an effective strategy for gaining knowledge from a SAM. 
        %They are interested in erroneous situations and demand communication in project situations. 
        %They
        They talk about how, in their assessment, a particular situation is a mistake in order to challenge that assessment through the perspective of others. 
        %This kind of communication is important but is not perceived as easy. 
        Talking about mistakes means a common evaluation of a situation, the comparison of ideas about an ideal situation, and thinking about further possibilities for action. 
        \textit{``Yes, mistakes need to be talked about.  %And communication is, I'm pretty sure, the more difficult and significant part than the solution. 
        \ldots It's only through communication and discussion that you come to a good alternative solution that you want to apply.''} 

        Secondly, whether architects are able to learn from SAMs depends on their personal openness to SAMs and the openness of the environment to SAMs. 
        On a personal level, we classified the attitude of all participants as constructive. 
        The participants were open to and aware of their own mistakes without seeing them as a defeat. %, and, therefore, they remained able to act. 
        Similar to the personal level, mutual acceptance of mistakes in the environment is also important
        %The question of guilt moves into the background. 
        The expectation should not be ``It is bad to make mistakes'' but ``It is bad not to learn from mistakes.'' 
        \textit{``We're all human, we all make mistakes, that's okay. 
        You have to accept that in order to get better. 
        Actually, it's not a mistake, but you learn something from it.''}  
        %Actually, the question is: How can we maximize making mistakes and maximize learning from them?''} 

        Thirdly, the participants stated that an established continuous improvement process (CIP)~\cite{imai_kaizen_1986} is essential for the learning process.  
        %A CIP is applicable on a personal and organizational level. 
        A CIP should specifically try to increase the triggers for SAM detection and make detection a regular process.  
        Furthermore, the SAM's detection is promoted by refining the idea of the ideal solution and the collection of information that can be used in the evaluation. 
        The CIP tries to derive concrete actions that prevent SAMs in the future. 
        \textit{``[After a SAM situation,] I directly asked: What could have been done better?  
        Could this have been foreseen beforehand? What is being done to avoid it in the future?''} 
        
        Finally, it is considered important to have the necessary information to classify the SAM situation. 
        For SAMs, the focus is on the context of the decision. 
        What were the reasons for the decision? 
        What were the goals? 
        %An appropriate assessment cannot be made independently of this information. 
        The documentation of decisions, e.g., by Architecture Decision Records (ADR)~\cite{van_heesch_documentation_2012}, is mentioned but not considered to be widespread.  
        \textit{``I am a big fan of Arc42 and also ADR \ldots 
        I often start a system design with quality scenarios and quality metrics, especially with new customers. \ldots 
        I also do a lot of architecture reviews.''} 
        %They say, `Can you do a review there?'. 
        %Then I say, `If you give me the quality metrics. What else am I doing this for? In terms of what do you want me to review?'} 

        \vspace{0,2cm}
        \noindent
        \fbox{%
            \parbox{0.97\textwidth}{
             Architects need to communicate the mistake with other stakeholders to be able to learn from SAMs.
             The ability to learn from SAMs depends on the openness to admit SAMs on a personal and organizational level. 
             An established CIP helps with SAM detection, communication, and information gathering.
             Important information regarding a SAM are SA decisions and their rationale, which should be documented. 
             }
        }

        \paragraph{\textbf{Preventing SAMs.}}
        
        Learning from mistakes is one form of preventing further mistakes.  
        \textit{``Even if it is a mistake, it is, of course, important, especially in the architecture, that you try to prevent it in other areas that might run into this same mistake.''}
        
        Furthermore, the architects approach people to challenge their decisions before implementing them. 
        \textit{``We had meetings, I presented the advantages and disadvantages from my point of view, and we did not get together afterward.''} 

        The architects also mentioned other means of preventing SAMs, e.g., taking responsibility for SA, making group decisions, and making more conscious decisions. 
        However, these topics are already discussed in detail in many other research papers.

        \vspace{0,2cm}
        \noindent
        \fbox{%
            \parbox{0.97\textwidth}{
             To prevent SAMs, we should communicate and learn from past SAMs.
             Challenging SA decisions, i.e., communicating them in advance, is a way of preventing SAMs.
             }
        }
        
        \subsection{RQ3: Communication of SA Mistakes}
        \label{sec:Results-RQ3}
        Communication is used as a strategy for SAM detection, for learning from SAMs, and for SAM prevention (see~\Cref{fig:SA_Mistake_Management}). 
        This section presents the communication partners, communication facilitating and hindering factors, and the structures in which communication takes place. 

        \paragraph{\textbf{Communication Partners.}}
        All interviewees work in agile project teams, and their most mentioned form of communication is communication on the team level, i.e., with team colleagues. 
        The colleagues are often directly affected by and interested in resolving the situation. 
        It is a common practice in their teams to discuss a solution in a group. 
        A special kind of group communication is discussions in the context of retrospectives. 
        \textit{``%At the project level, 
        This has led to retrospectives in which people have said: 
        `That's annoying, and it keeps coming back. Can we do something about it or not?' 
        %That happened before this decision [realization] `Maybe it was wrong.' had set in. 
        %At some point, we said: Why is this happening?
         \ldots
        Then we had a special meeting about it with the topic: Where are we? Where do we stand? Where do we have to go?''} 

        The client is also a frequent target of communication, especially due to the need to coordinate further action and the financial impact of decisions. 
        Involving the client in retrospectives or debriefings is perceived as valuable. 
        Likewise, in acute SAM situations, the client is usually involved. 
        In such a situation, some architects considered communication to be more difficult in some companies. 
        \textit{``
        %If you only have internal colleagues, that is, you can really talk openly about it [SAMs] \ldots 
        [I]f you point out mistakes in a project team with other external people or even with customer employees, particularly in a customer environment where you also have other [consulting] companies, you really have to be very sensitive and really think twice about what you say.''} 

        In the context of SA decisions, communication with direct superiors was not considered to be difficult, e.g., \textit{``I would assess myself in such a way that I can admit mistakes, even to my boss.''} 
        One participant even stated:
        \textit{``My boss gets to listen to a lot. And my boss's boss and my board. So, I'm really there -- I have a big mouth. I so don't care.''}

        The participants rarely mentioned talking about SAMs with architects outside the team.
        This is only done if necessary by asking specific questions or in the context of an architecture board. 
        \textit{``[T]here is a regular meeting once a month, called the architecture circle. 
        %When we make important architectural decisions in the project, I say: `It's not enough to just think about it here. 
        %We  \ldots throw that on the table at the architecture board, describe the problem we want to solve, and ask: `What ideas do you have?' 
        And then all the brains of my employer sit there and give their suggestions.''} 
        
        Concrete examples of active company-wide communication of SAMs are scarce.  
        Special formats for the presentation of mistakes, such as `Fuckup Nights,' are mentioned only once.
        Open spaces and all-hands workshops are also mentioned as spaces with the possibility that SAMs are discussed. 
        \textit{``We do a lot of `Open Spaces,' including technical `Open Spaces.' Half of them are not about how cool something is that you built but to look at the mistakes I made.''}

        Very occasionally, SAMs are discussed with outsiders, e.g., in the form of conference talks or networking events.  
        The personal conviction that SAMs present valuable knowledge is the motivation for this. 
        \textit{``I think I can teach people more when I talk about [SA] mistakes than how to do it right.''}

        \vspace{0,2cm}
        \noindent
        \fbox{%
            \parbox{0.97\textwidth}{%
             SAM communication takes place on six different levels: with team colleagues, with the clients, with the superiors, with other architects, company-wide, and with outsiders. 
             The most common form is communication within the team.  
             Company-wide discussions of SAMs or communication with outsiders rarely occur. 
             }
        }
        
        \paragraph{\textbf{Communication Facilitating and Hindering Factors.}}
        The communication preferences and observed communication patterns can be divided into facilitating and hindering factors.  
        However, some of these factors can be seen as both facilitating and hindering, depending on the architects' characteristics, e.g., professional status and self-confidence, and psychological safety.

        The biggest facilitating factor in communicating SAMs successfully is ensuring access to information. 
        This access to information is influenced by the diversity of experiences, i.e., multiple cultural backgrounds but also different domains of IT experience. 
        Diversity ensures that many requirements are represented.  
        \textit{``We have a regular meeting in development. \ldots
        Anyone can bring in any topic, be it from testers or other teams. 
        Not development teams, but other teams. 
        That's when a few problems come to light.''} %where you say: `Yes, interesting, let's talk about this in a concentrated group and find a solution.'''} 

        However, this also means a diversity of communication preferences. 
        Responding to these preferences or even being aware of differences between team cultures can further facilitate communication.
        \textit{``I am also someone who is not afraid of speaking in large crowds. 
        There are also people who perhaps have relevant experience who don't want to speak in front of a large crowd right away, where [for whom] you would then perhaps have to find other means to communicate that.''} 

        Transparent communication promotes exchange. % because it gives other people a way to respond to the information. 
        Having access to or being exposed to SAMs information enables the absorption of information, which means that learning from the mistakes of others is possible. 
        In the office, you inevitably run into other people. 
        In times of remote work, teams need to communicate in open channels rather than in private. 
        \textit{``Otherwise, I often ghost around [roam around] in some channels, i.e., Microsoft Teams channels, in order to also capture topics that are being discussed.''}
        %The POs, i.e., the product owners, hold a lot of meetings. So day workshops, every week. I don't participate in those. \ldots 
        %But I try to then skim the extract from it.''} 

        The biggest hindering factors are potential negative consequences. 
        The communication of SAMs can generate costs, e.g., in contracting situations or raise the question of guilt.  
        \textit{``Unfortunately, it is often the case that software architecture is the most expensive decision and that mistakes are often not communicable.''}
        %Some participants reported negative experiences with colleagues who relayed this pressure to others.  
        Additionally, the personal expectation of the outcome of communication can trigger negative consequences in individuals, such as disappointment. 
        \textit{``With younger colleagues, the disappointment is usually great when this [developed SA proposal] decision is not followed.''} %  Older ones are more like me and say, `You can't do more than suggest it to the customer.'''} 

        The participants also consider the environment to be of decisive importance for the willingness to communicate SAMs. 
        \textit{``If you're in the corporate bubble, then most people \ldots can admit mistakes and reflect on them. In other environments, I have experienced different situations where people can't admit it [mistakes].''}% or don't even have a basis for conversation.''} 
        % %CUT:
        % The architects emphasize that there is no concern for dismissal due to faulty decisions. 
        % However, it is perceived positively when others confirm that communicating SAMs is desirable or that pointing out mistakes through feedback is considered valuable. 
        % The preference to work in such environments is also clearly stated. 
        % %bis hier
        This is called psychological safety in the domain of psychology, which promotes information sharing~\cite{newman_psychological_2017}. 
        \textit{``[My] Superiors always say that reflection is important; admitting mistakes is important.''} 
        A lack of psychological safety hinders SAM communication. 
        
        Furthermore, discrepancies among communication participants in reputation, status, or hierarchy seem to make communication more difficult. 
        \textit{``That was a challenge, to do a [meeting] format with a developer from a customer and his board of directors%at the same time
        .''}         
        % It was good that we had people who could moderate and had experience with it. 
        % Otherwise, a developer simply doesn't say anything critical in that context.''} 

        The participants also mentioned that as one's own professional status and self-confidence grow, the willingness to communicate SAMs increases, particularly in unclear situations. 
        This, however, means that stakeholders with less status and self-con\-fi\-dence might not share their experiences with mistakes. 
        \textit{``So that you can admit mistakes openly, you also need a certain standing. 
        And if you don't feel confident, you won't do it.''} 
        % Especially if it is not objectively clear that there are mistakes. 
        % Then you can avoid the discussion.''} 

        People do not only act rationally, either. 
        The exchange of SAM information is hindered by human characteristics such as emotional bias, e.g., human conflicts may hinder communication. 
        Seeking direct conversation does not work if there is no access or willingness to talk. 
        % \textit{``Actually, the mistake is small. 
        % As management, you probably don't see it that way. 
        % There [That] was probably just [a case of], `I'm not really listening to my people.' 
        \textit{``So the question is, as an employee, how do I get the board to listen to me? 
        That's very difficult.''} 
        
        Finally, communication is complicated by conflicts of interest between the role in the project context and personal goals.  
        Even if there is no fear of dismissal, communicating SAMs could damage one's own reputation or that of the company, no further assignment may follow, or conflicts may arise. 
        %Due to the different goals of different roles in the project context, the communication of SAM's information can be contrary to the personal maximization of goals. 
        \textit{``But sometimes that is not the right architecture at all.  
        Nevertheless, you might defend it that way. 
        You're not going to admit that mistake and say, 'I did that because it helps me sell my people [consultants].''}

        \vspace{0,2cm}
        \noindent
        \fbox{%
            \parbox{0.97\textwidth}{%
             Facilitating factors for SAM communication are access to information, i.e., transparent communication and diverse teams in terms of cultures and expertise, as well as supportive communication structures.
             Hindering factors are potential negative consequences, a missing psychological safety, discrepancies in reputation, status, or hierarchy, a missing professional status and self-confidence, a missing psychological safety in the environment, emotional biases, and conflicts of interest. 
             }
        }
        
        \paragraph{\textbf{Communication Structures and Strategies.}}

        The participants' preferred communication structure is to talk directly with others. 
        \textit{``%If a question is simple at first, then I simply approach them [colleagues]. 
        In a team, we are closer to each other, and we sit down together when there are issues.''} 
        
        While the participants seek direct communication for acute problems, regular appointments are nevertheless a frequently observed pattern for active exchange.  
        These appointments take place regularly and for various purposes.
        They shape communication not just in response to problems but proactively: 
        \textit{``These classic escalation issues have become fewer because we have managed to have regular appointments. 
        As overarching architects, we have `Face To Faces' with all the teams  \ldots to sit down and identify and discuss issues.''} 

        During these conversations, participants pursue the strategy of communicating constructively and explaining their own points of view in a well-founded manner. 
        In this context, the participants perceive it as helpful not only to communicate problems and not to create accusations. 
        \textit{``You have to say, `I don't think this is well designed, for such and such reasons, we have such and such disadvantages.'  \ldots It's not about, `Peter, you wanted to do it then, and this is bad now.''}
        %This constructive clarification of what's wrong is important, of course. ''} 
        
        Predominantly, architects try to convince their colleagues rather than overrule them.  
        By this, they hope that the other person will adopt the values in the long term rather than simply accept them without understanding them. 
        \textit{``So they might do it then, grudgingly accept it. But they do not understand it. And without understanding, they won't do it any better next time.''}

        \vspace{0,2cm}
        \noindent
        \fbox{%
            \parbox{0.97\textwidth}{%
            Architects prefer direct communication with colleagues. 
            Regular appointments are important and have the advantage of addressing problems proactively.
            Communication should be constructive and personal positions should be explained in a well-founded manner. 
            It is important to convince people rather than overrule them. 
             }
        }

    \section{Discussion}
    \label{sec:Discussion}
    %In this section, we discuss the results of ~\Cref{sec:Results} with regards to  .

    In retrospect, the initial hypothesis that SAMs are typically not sufficiently communicated has been only partially confirmed. 
    In the project teams and the peer groups of the participants, there is an active exchange about SAMs.
    %Whether SAMs are communicated is largely attributed to the environment, which is strongly influenced by the involved persons. 
    %They demand communication and try to establish a corresponding process. 
    %Based on the assumption that the sample is biased toward individuals who value communicating SAMs, this seems to be driven by personal motivation. 
    %At the same time, individuals report that others give them feedback that their approach is valuable. 
    However, the situation changes on the company level or outside the company. % regarding the question of whether this knowledge is also imparted across the company. 
    Despite a high level of commitment on the part of the participants, there are only a few examples in which SAMs are discussed outside the involved project teams. 
    In terms of the participants' companies' size, this means that many employees do not have access to the learned information. 
    
    Obviously, the processing of information in a larger circle must be approached differently and requires more effort than in a small circle of individuals. 
    The information and facts must be prepared more systematically, and there is less time for individual questions. 
    However, making an information exchange process publicly accessible within the company, e.g., in open spaces or all-hands workshops, would counteract information silos. 

    % Learning thus takes place predominantly in project teams and peer groups of individuals. 
    % In this context, the demands on the communication skills of software architects with respect to SAMs are high.  
    % This is particularly the case for situations with a high degree of conflicts of interest or status differences.  

    Practical indications can be derived from the action strategies of the participants. 
    Regarding the process model presented in \Cref{fig:SA_Mistake_Management}, goals can be formulated, which are discussed below.
    % \begin{itemize}
    %     \item Increase the number of triggers for SAM assessment
    %     \item Ensure communication of SAMs
    % \end{itemize}

        \subsection{Increase Detection Trigger} 
        People and organizations can influence the triggers for SAM detection in the process shown in \Cref{fig:SA_Mistake_Management}. 
        More triggers mean more opportunities for the SAMs' detection. 
        
        On a personal level, a constructive mindset is most crucial. 
        A different personal attitude may cause SAM detection to be blocked. 
        The point perceived as most important is the change in expectations mentioned in \Cref{sec:Results-RQ2}. 
        ``It is bad to make mistakes'' changes to ``It is bad not to learn from mistakes.'' 

        On a personal and organizational level, formalized and regular CIPs should be established. 
        %Such a process should be characterized by a certain formalization and regularity. 
        They can be used to check the validity of SA ideas and implementations and is thus a trigger for SAM detection. 
        %With regard to architectural decisions, there is also a focus on evaluation. 
        The participants report regular architecture reviews or exchanges with architects as one way to achieve continuous improvement, but they require lightweight approaches. 
        %They focus on a reduced level of effort and content compared to larger assessment procedures such as ATAM by focusing on the big picture and regular execution every 2--4~weeks.  

% %CUT:
%         %Lack of knowledge was a key factor in the causes of SAMs. 
%         More knowledge about SA in agile project teams presumably not only leads to better decisions but to introspective triggers occurring more frequently. 
%         Architectural knowledge must be present in teams that have to make architectural decisions. 
%         Recommendations from the participants are to join peer groups, i.e., ``architecture communities'' like Meetups~\cite{noauthor_meetup_2023}. 

        \subsection{Ensure Communication Opportunities} 
        
        Our results show that it is extremely important that SAMs are communicated. 
        Successful communication involves not only talking but also listening. 
        The stigma of failure must be removed so that open discussion, rapid learning, and innovation can occur instead of employees hiding failures in order to protect themselves. 
        Conditions for open discussion should be created, and invitations to participate should be formulated and responded to productively. 

        The negative connotation of SAMs was discussed in \Cref{sec:Results-RQ1}. 
        Therefore, some people try to portray SAMs in a positive or euphemistic way. 
        A constructive mindset also means learning from mistakes. 
        However, this should not lead to a euphemistic presentation of SAMs. 
        SAMs, in general, are not positive but have negative consequences and are avoidable. 
        The fact that they also have positive consequences does not make them positive as a whole. 
        
        Firstly, these findings can be taken as an impetus for an analysis of one's own handling of SAMs.
        Secondly, they are also a call for companies to further enable and support SAMs communication among their employees, e.g., by creating company-wide communities of interest. 
        Finally, there should be more open communication about SAMs outside of companies or across companies. 
        For example, governments or chambers of commerce could support this information exchange across companies.

    \section{Related Work}
    \label{sec:RelatedWork}
        Our related work section focuses on the main topics of our work, which are SAMs and their communication.
        
        \paragraph{\textbf{Software Architecture Mistakes.}} 
        The topic of SAMs is highly related to SA decision-making. 
        A variety of research has been done in this research field, e.g., ~\cite{Bhat2020}. 
        However, most of this research focuses on the decision-making process when deciding on an SA or a part of a SA. 
        
        Tang et al.\ did a systematic literature review of the human aspect in decision-making~\cite{tang_human_2017}. %, which is related to our work as the human aspects are essential to our research, too\cite{tang_human_2017}.
        One of the mentioned decision-making practices is ``knowledge management'', i.e., knowledge capture, sharing, and communication, which are also important aspects of our work.
        Yet, all mentioned studies focus on knowledge documentation.
        While Tang et al.\ acknowledge that ``experience and knowledge play a role in decision making,'' all identified studies focus on the knowledge aspect~\cite{tang_human_2017}.
        The experience aspect, which includes learning from other people's experiences, is not mentioned.
        In their semi-systematic literature review, Bhat et al.~\cite{Bhat2020} also identified different studies stating the relevance of experience and knowledge, e.g.,~\cite{hassard_analogies_2009}.  %~\cite{hutchison_foundations_2006} or
        Yet, no work seems to explore how to improve the experience of architects.
        
        Less research has been done on how to deal with bad decisions after they have been made. 
        Some of these aspects are discussed in the research field of architectural technical debt, e.g.,~\cite{Besker2018a}.  
        For example, Kruchten et al. present a guideline on how asking about SAMs might lead to the root causes for some architectural technical debts~\cite{Kruchten2019}.
        However, in their recent interview study, %of the IT managers' perspective on technical debt
        Wiese et al.\ still identified a lack of refactoring strategies for old legacy systems as IT managers' main concern~\cite{wiese_it_2023}. 

        \paragraph{\textbf{Software Architecture Mistake Communication.}}
        In the field of SA Communication, a wide range of work has been done, e.g., regarding documentation~\cite{van_heesch_documentation_2012}, visualization~\cite{shahin_systematic_2014}, or stakeholder communication~\cite{marquez_involving_2021}. % (Taušan,) MÁRQUEZ. %CUT candidate: Tausan nehmen, weil Title von Marquez sehr lang
        Smolander et al.\ argue the importance of organizational and inter-organizational communication between architects~\cite{schmolander_describing_2002}. 
        However, regarding the communication of SAMs between architects, we were not able to find related studies. 

        We argue that communication is essential to amass SA knowledge.
        In their study about SA knowledge in search engines, i.e., Google, Soliman et al.\ identified a lack of knowledge on bad decisions in the search engine results~\cite{Soliman2021c}. 
        Our study supports the idea that this lack is not caused by a faulty search strategy but by a lack of willingness to share this knowledge.
        However, we did not find research on if or how communication of SAMs might enhance knowledge gathering.

    \section{Threats to Validity}
    \label{sec:ThreatsToValidity}
        We present threats to validity for our qualitative study by addressing the construct, internal and external validity. %based on the guidelines provided by Wohlin et al.~\cite{Wohlin2012}.
    	
    	\paragraph{\textbf{Construct Validity.}}
    	%Construct validity refers to the appropriateness of the chosen methods.
            For an interview study, the selection of participants and interview questions have a great impact on the validity of the study. 
            In our study, we only interviewed architects working in an agile environment. 
            As we did not enforce this as a precondition, we suspect that an agile team organization is state-of-the-art. 
            We, therefore, suggest using our study's results specifically for agile working companies. 
            %However, we did not particularly search for participants from other organizational environments and cannot exclude any influence. 

            Furthermore, architects that are more open to discussing their SAMs might be more willing to participate in our study, which might bias our study. 
            We also recruited participants from our personal network.  
            These participants might not be as open regarding their SAMs but might still participate for our sake. 

            Finally, we interviewed ten architects, which might seem too few. 
            However, we reached code saturation on the focus code level, i.e., codes that were reported in this paper did not change during the last three interviews. 
            Further studies in other countries and with more participants, e.g., surveys, should validate our theories.
            
    	\paragraph{\textbf{Internal Validity.}}
             %Internal validity threats are related to possible wrong conclusions about causal relationships between treatment and outcome. 
            In the case of our study, internal validity means the possibility of wrongly interpreting and misunderstanding the interview transcripts. %, e.g., by misunderstanding the participants.
            To avoid such misconceptions, the coding was done by two researchers. 
            Furthermore, we were able to ask the participants for clarification in cases we were unsure about a statement. 
    	
    	\paragraph{\textbf{External Validity.}}
    	%External validity threats refer to the ability to generalize the result.

            External validity concerns the generalization of the result. 
            However, the ability to generalize the results is not the goal of a qualitative study such as ours. 
            Nonetheless, we only surveyed German architects, and architects in other countries might have different insights, particularly as cultural aspects are important in our study.
            The generalizability of our results could be enhanced by further studies, e.g., by surveys generating quantitative data and by interview studies in other countries.
    
    	%\paragraph{\textbf{Conclusion Validity}
    	%Conclusion validity relates to the reliability of the conclusions drawn from the results.    

\section{Conclusion}
\label{sec:Conclusion}
        In our study, we analyzed the communication of SAMs in the industry by performing an interview study with ten architects. 
        We derived their definition and list of characteristics of SAMs, their handling of such SAMs after their detection, and finally, their ways of communicating SAMs. 
        We identified optimization potential regarding the enhancement of SAM's detection triggers.
        
        While the communication of SAMs takes place on the team level, we identified optimization potential on a company and cross-company level. 
        Furthermore, we identified communication partners, structures, and strategies, and we presented communication facilitating and hindering factors.

        \textbf{For practitioners,} our study could spark the company-wide exchange of SAMs. 
        %IT managers could launch and support initiatives to establish enterprise-level exchange meetings. 
        Furthermore, we provided many ideas on how to optimize communication on a personal level, e.g., by visiting public communities of interest and changing our own mindset from ``It is bad to make mistakes'' to ``It is bad not to learn from mistakes.''

        \textbf{For researchers,} we provided the theory of model on how practitioners manage SAMs. 
        This, in turn, forms the question of how to decide whether to accept a SAM, change the architecture, or revise the architecture completely. 
        Furthermore, we identified the research question on how to optimize the communication of SAMs on the enterprise level besides the ideas mentioned by our participants.  
        For these ideas, an evaluation of their effects is also still missing. 

        \textbf{For future work,} we suggest performing a subsequent survey study to validate our theories and refine them by taking the different contexts, e.g., domains, countries, and cultural environments, into account. 
        In addition, this could help to generalize our results. 
        %Furthermore, it would be intriguing to replicate this study in other countries and cultural environments.
        After that, the aforementioned research questions on how to optimize SAM communication should be addressed by solution-seeking studies.

% \section*{Acknowledgment}

%\section*{References}

\newpage 

\bibliography{ArchitectureMistakes}





% \end{thebibliography}
\end{document}